\documentclass[twocolumn]{pasj00}

\begin{document}

\SetRunningHead{Suzaku Results on the Obscured Low-Luminosity Active Galactic Nucleus in NGC~4258}{S. Yamada et al.}
\title{Suzaku Results on the Obscured Low-Luminosity \\ Active Galactic Nucleus in NGC~4258}

\author{Shin'ya \textsc{Yamada},\altaffilmark{1}
Takeshi \textsc{Itoh},\altaffilmark{1}
Kazuo \textsc{Makishima},\altaffilmark{1,2}
Kazuhiro \textsc{Nakazawa},\altaffilmark{1}
}
\altaffiltext{1}{
   Department of Physics, University of Tokyo\\
   7-3-1, Hongo, Bunkyo-ku, Tokyo, 113-0033, Japan}
\altaffiltext{2}{
   Cosmic Radiation Laboratory, Institute of Physical and Chemical 
   Research (RIKEN)\\
   2-1 Hirosawa, Wako-shi, Saitama, 351-0198, Japan}

\email{yamada@amalthea.phys.s.u-tokyo.ac.jp}


\Received{$\langle$reception date$\rangle$}
\Accepted{$\langle$accepted date$\rangle$}
\Published{$\langle$publication date$\rangle$}

\maketitle

\begin{abstract}
In 2006 June, the obscured low luminosity active galactic nucleus in the nearby Seyfert 1.9 galaxy NGC~4258 was observed with Suzaku for $\sim$ 100 ks.
Utilizing the XIS and the HXD, 
the nucleus emission was detected over a $\sim$ 2 to $\sim$ 40 keV range,
 with an unabsorbed 2--10 keV luminosity of $\sim 8 \times 10^{40}$~erg~s$^{-1}$, and varied by a factor of $\sim$ 2 during the observation. 
Its 2--40 keV spectrum is reproduced 
by a single power law with photon index of $\Gamma$~$\sim$ 2.0, 
absorbed by an equivalent hydrogen column of $\sim$ 1.0 $\times$ $10^{23}$ cm$^2$. 
The spectrum within $4'$ of the nucleus required also a softer thin-thermal emission, 
as well as an intermediate hardness component attributable to integrated point sources.  
A weak neutral Fe-K$\alpha$ florescence line was detected at an equivalent width of $\sim 40$ eV. 
The cold reflection component was not required by the data, with the reflector solid angle $\Omega$ seen from the nucleus 
constrained as $\Omega / 2 \pi$ $\lesssim 0.3$ assuming a general case of 60$^\circ$ inclination.
The results suggest that the cold reflecting material around the nucleus is localized along our line of sight, rather than forming a thick torus.
\end{abstract}

\section{Introduction}
\label{sec:intro}
Although active galactic nuclei (AGNs) exhibit a wide variety of spectral properties, 
a comprehensive classification of them, called ``Unified Scheme'' (e.g., \cite{Antonucci1993}), has been developed. 
In addition to the black hole mass and the mass accretion rate which are obvious parameters, 
this scheme employs two more fundamental parameters: radio loudness and our viewing direction to the accretion plane.
A ``type~1'' AGN refers to an object with a nearly pole-on viewing angle,  
while a ``type 2'' AGN with a roughly edge-on aspect. 
The spectrum differs between the two classes depending on whether 
our line of sight is blocked by some part of the accreting material or not.  
The blocking material is often assumed to have a shape of an inflated torus around the nucleus, so called ``molecular torus'' or 
simply ``torus". 
This scheme can explain much of the observational differences
among various kinds of AGNs. Nevertheless, the validity of the Unified Scheme and the nature of the putative torus are 
still open.

X-ray observations of type~2 AGNs provide a powerful tool in these attempts, 
because the strength of photoelectric absorption affecting the nuclear hard X-rays provides the most direct information on the column 
density of the obscuring material. Indeed, X-ray observations with Ginga (e.g., Koyama et al. 1989; Awaki et al. 1991), and ASCA ( Ueno et al. 1996) 
of typical type~2 AGNs demonstrated that their nuclei are heavily obscured with a column density of $10^{23}$ to $10^{24}$ cm$^{-2}$. 
Subsequently, BeppoSAX investigators introduced a new concept called ``Compton thick AGNs" (\cite{Maiolino1998}, \cite{Matt2000}),
 as the most extreme class of type 2 AGNs, wherein the absorber is opaque not only to photoelectric absorption, but 
 also to Compton scattering. Because of the strong suppression of the direct component, these objects are expected to provide further 
 clues to the circumnuclear matter distribution, via detection of various secondary components, including in particular  
 the Compton reflection hump, and fluorescent Fe-K lines.

Using Suzaku, we observed the Compton-thick AGN, NGC~4945, which is one of the brightest AGNs above 20 keV.
Analyzing the data, Itoh et al. (2008) found the reflection features to be unusually weak, as judged from both
the Fe-K edge feature in the XIS spectra and 
the Compton hump in the data from the HXD. 
Reinforced by the detection of clear hard X-ray variations, Itoh et al. (2008) 
thus constrained the reflection fraction \ensuremath{f_\mathrm{refl}} $\equiv$ $\Omega$/2$\pi$ to be less than
 a few percent, where $\Omega$ is the solid angle of the reflector as seen from the nucleus.  
This makes a significant contrast to many other Compton-thick Seyfert 2 AGNs, where we usually find 
\ensuremath{f_\mathrm{refl}} $\sim$ 1-2 (e.g. De Rosa et al. 2008). 
Therefore, the reflector in NGC~4945 is suggested to have a geometrically-thin, disk-like structure 
rather than that of a thick torus, and the reflector, absorber, and the water maser source are thought to be provided by the same structure, 
namely a flat accretion disk (Madejski et al. 2000, 2006). 
Although this means a clear deviation from the Unified Scheme, it is not yet clear whether such a property is 
specific to NGC~4945, or more or less common to a certain class of AGNs. 
One of effective ways to examine this issue is to study AGNs with low mass-accretion rates, namely 
low luminosity AGNs (LLAGNs), because NGC~4945 has a rather low intrinsic (absorption corrected) luminosity as 
$\sim 1 \times 10^{43}$erg s$^{-1}$ in 2--10 keV. 

NGC~4258 (M~106), one of the typical LLAGNs, is a highly inclined
(72\degree, \cite{{Tully1988}}) SABbc spiral galaxy located at a nearby 
distance of 7.2~Mpc (\cite{{Herrn1999}}), where $1''$ corresponds to 35~pc. 
The presence of an obscured AGN was suggested by the strong polarization of the relatively broad optical emission lines
(\cite{Wilkes1995}), and H$\alpha$ and X-ray emitting ``anomalous arms" (Cecil et al. 1995). 
The observation with ASCA up to 10 keV (Makishima et al. 1994) clearly demonstrated the presence of an LLAGN, 
absorbed by an equivalent hydrogen column of $N_\mathrm{H} \sim 1.5 \times 10^{23}$~cm$^{-2}$, with an absorption-corrected  
2--10~keV luminosity of $\sim 4 \times 10^{40}$~erg~s$^{-1}$. 
The obscured nucleus was later reconfirmed by a Chandra image (\cite{Wilson2001}; \cite{Yang2007}). 
The epoch-making water maser observation (\cite{Miyoshi1995};
\cite{Greenhill1995a};
\cite{Greenhill1995b}) established the AGN mass as $3.6 \times 10^7$~\ensuremath{M_\odot}, and its disk inclination as $\sim 83\degree$. 
The X-ray flux is known to be variable on time scales of hours to years (\cite{Reynolds2000}; \cite{Terashima 2002}; 
\cite{Fruscione2005}; \cite{Fiore2001}). 
According to observations with XMM-Newton (Pietsch and Read 2002) and Chandra (Yang et al. 2007), 
contributions by other softer X-ray components associated with NGC~4258 can be spectrally separated from
the harder nuclear emission.
Among them, the soft emission extending out from the nucleus, along ``anomalous arms'' is of particular interest, 
because of its possible association with the AGN jets
(e.g., \cite{Wilson2001}), but it is beyond the scope of the present paper.

The most outstanding property of the NGC~4258 nucleus is its extremely low luminosity, not only in the absolute sense                                                          
but also relative to the Eddington value. Indeed, its absorption-corrected 2--10 keV luminosity is only                                                                                                                               
$\sim 8 \times 10^{-6}$ of the Eddington value, 
which is much lower than those of Seyfert galaxies                      
(typically thought to be of the order of $\sim$ 1\%).
Therefore, its accretion flow could be significantly different from those of Seyferts; 
e.g., the concept of Advection Dominated Accretion Flow 
(ADAF; Ichimaru 1977; Narayan and Yi 1994, 1995) may be more applicable 
to these LLAGNs than to Seyfert galaxies. Nevertheless, broad-band X-ray properties of NGC~4258 
are not much different from those of radio-quiet Seyferts, 
as reported by Fiore et al. (2001) based on a BeppoSAX detection up to $\sim 70$ keV 
with a loosely constrained photon index of $\sim 2.1$. 
Then, we may need to look for more subtle differences, e.g. in the reflection hump and Fe-K lines.
Obviously, Suzaku is most suited for this purpose. 
In the present paper, we report on a tight upper limit on the reflection 
component in this LLAGN, and spectral changes on a 
time scale of $\sim$ 50 ks, both derived from a Suzaku observation made in 2006 June.  

\section{Observation and Data Processing}
\label{sec:obs} 
\subsection{Observation} 
\label{subsec:obs} 
During the AO-1 cycle, we observed NGC~4258 with Suzaku for a gross time span of 187.5~ks, 
from UT 12:49 on 2006 June 10 through UT 17:04 on June 12. 
The observation was carried out with the source placed 
at the center of the HXD field of view. 

The XIS (Koyama et al. 2007) and the HXD (Takahashi et al 2007; Kokubun et al. 2007) were 
operated both in the nominal modes throughout the observation. 
Event files from the two instruments were screened using version 2.2.7.18 
of the Suzaku pipeline processing. We used ``cleaned events'' files, in which data of the following criteria were discarded:
data taken with low data rate, or with an elevation angle less than 5$^\circ$ above dark Earth, 
or with elevation angles less than 20$^\circ$ above sunlit Earth, or during passages through or close to the South Atlantic
Anomaly (SAA). Cut-off rigidity (COR) criteria of  $>$ 6 GV were applied to 
both the XIS and HXD data. 
In the present observation, the source was detected significantly with the XIS (section 2.2), 
and HXD-PIN (section 2.3), but not with HXD-GSO. 

\subsection{XIS Data Selection and Background Subtraction
\label{subsec:obs_xis}} 

XIS events with grades 0, 2, 3, 4 and 6 were selected, and then hot and flickering 
pixels in each CCD chip were removed using \texttt{cleansis} software. 
Using \texttt{aeattcor} software (Uchiyama et al. 2007), we corrected the event positions for attitude fluctuations
due to thermal spacecraft wobbling. Then, we extracted those events which satisfy the criteria described in section 2.1. 
A net exposure of $\sim 99$~ks was obtained in total from each XIS sensor. 

\begin{figure}
 \begin{minipage}{.48\textwidth}
  \begin{center}
  \FigureFile(80mm,80mm){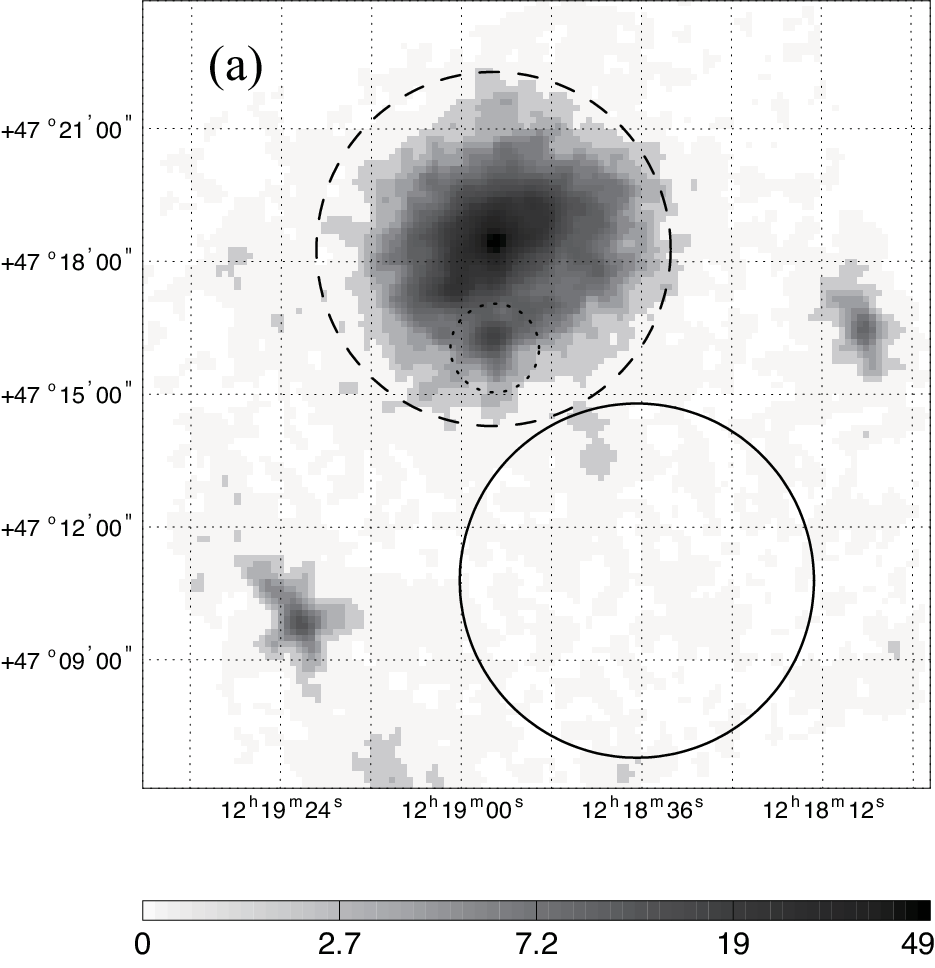}
  \end{center}
 \end{minipage}
 \hfill
 \begin{minipage}{0.48\textwidth}
  \begin{center}
    \FigureFile(80mm,80mm){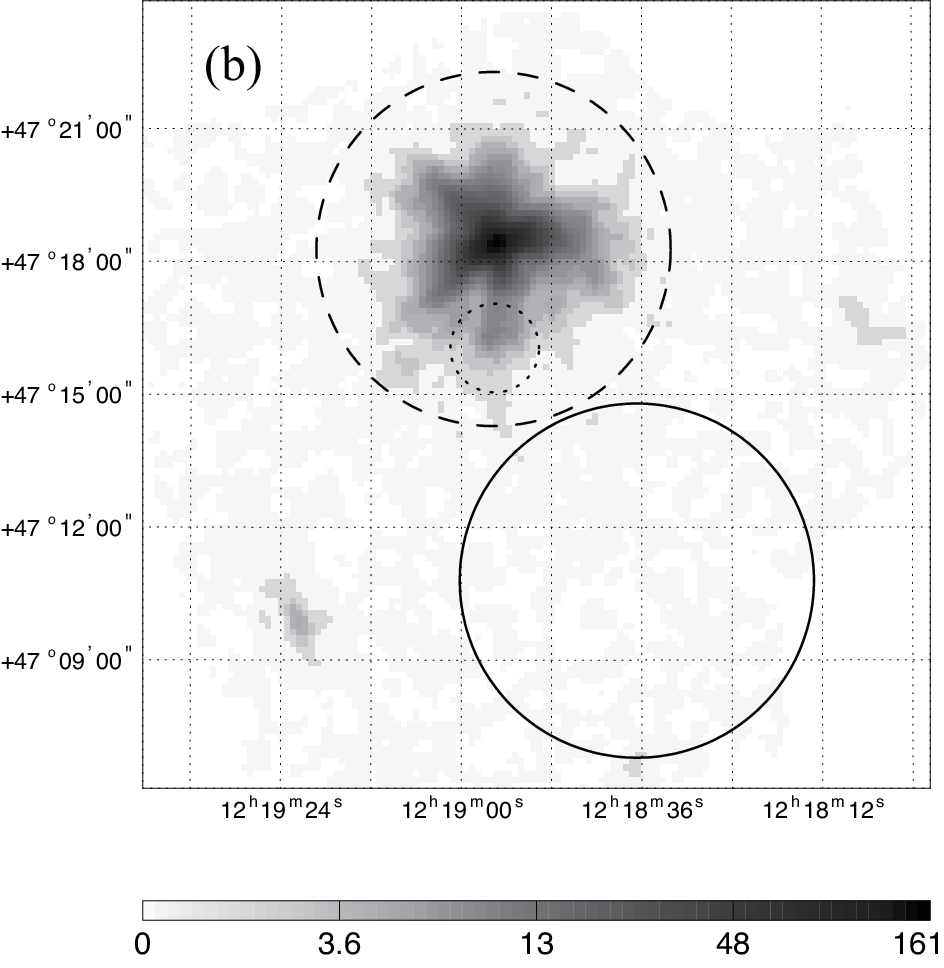}
  \end{center}
 \end{minipage}
 \caption{Background-inclusive Suzaku XIS0 images
 of NGC~4258, in the 0.3--3.0~keV (panel a) and 3--10.0~keV (panel b) band.
Dashed and solid circles, both of radius $4'$, indicate the on-source and background accumulation regions. 
A dotted small circle of radius $1'$ is a region used to eliminate a possible point source.}
 \label{n4258-img} 
\end{figure} 
 
Figure 1 shows XIS0 images of the NGC~4258 region, in two energy ranges
of 0.3--3~keV and 3--10~keV. 
The soft-band image is dominated by complex extended emission with an extent of at least $\sim 4'$,
while the hard-band image is dominated by the bright nucleus as first revealed with ASCA (Makishima et al. 1994).
In addition, several fainter point-like sources are
distributed over the NGC~4258 disk. 
As indicated in figure 1, we accumulated XIS events from a circular region of radius $4'0$ (8.4 kpc at the distance of 7.2 Mpc) 
centered on the nucleus. 
Since the nuclear emission is confused in this region with the extended galaxy emission, 
we restrict the following spectral studies mainly to energies above 1.9~keV.

At $\sim 2'.2$ south of the nucleus, 
figure 2 reveals a faint point source which has not been reported previously. 
We excluded a circular region centered on the source,
with a radius of $1'.0$ 
(indicated by a black dashed circle). 
Besides, the $4'$-radius region for the nuclear signal accumulation contains a Chandra point source, J121857.3+471812 (Yang et al. 2007), 
located at $\sim 1''.5$ off the nucleus, of which the contribution is estimated to be $\sim$ 3 \% at 3 keV and $\sim$ 1 \% 
at 5 keV of the nucleus. We consider its contribution in section 3.1, 
together with those from other fainter point sources that can contaminate own spectra. 

The background spectra were extracted from
a source free region of the same field of view, as indicated by a black solid circle in figure 1. 
The background count rates in 3--9 keV were found to be $\sim$7\% of the total event rate in the same energy range 
detected with each FI CCD from the on-source region, and $\sim$10\% of those with the BI CCD.
The response matrices and ancillary response files
were created utilizing \texttt{xisrmfgen} and \texttt{xissimarfgen} (\cite{Ishisaki2007}), respectively. 
In the spectral analysis described below, we coadded
the spectra and responses of the three FI chips
(XIS 0, 2, and 3). 
 
\subsection{HXD Data Selection and Background Subtraction
\label{subsec:obs_hxd}}
\begin{table}
 \caption{The HXD count rates during the Earth occultation periods, compared with the NXB model.}
 \label{earth4258}
 \begin{center}
  \begin{tabular}{llrr}
   \hline\hline
   Component & 12--40 keV & 40--60 keV \\
   \hline
   Earth event      & 0.4367 $\pm$ 0.0069    
                       & 0.0492 $\pm$ 0.0023 \\
                       
   NXB$^*$        & 0.4504 $\pm$ 0.0021
                       & 0.0464 $\pm$ 0.0007 \\[1.5ex]

   ratio      & 1.03 $\pm$ 0.02 
                       & 0.94 $\pm$ 0.05 \\
      \hline\hline
\end{tabular}
\end{center}
\begin{itemize}
\item[$^*$] Modeled with LCFITDT \citep{Fuka2008}.
\end{itemize}
\end{table}

Table 1 summarizes the count rate of HXD-PIN averaged over the whole observation. 
To subtract the non X-ray background (NXB), 
we used an NXB model, called \texttt{LCFITDT} or ``tuned'' method \citep{Fuka2008}, which accurately 
takes into account various variations of the NXB. 
The model also includes slight differences caused by the different bias voltages imposed on the 64 HXD-PIN diodes. 
This is because a quarter of the whole HXD-PIN diodes 
have been operated with a bias voltage of 400~V since 2006 May 24 (to suppress anomalous noise behavior seen in some of the diodes) 
while the others were operated at 500~V. 
Systematic errors of the NXB model in 15--40 keV and 40--70 keV are estimated as 1.4\% and 2.8\% (1 $\sigma$ level), 
respectively, when night earth is observed for an exposure of 10 ks (Fukazawa et al. 2008). 
Since our observation is longer than 10 ks, the NXB model is expected to have a reproducibility no worse than these estimates.  
Therefore, we use these values as an approximation to the systematic errors involved in the NXB subtraction.

\begin{figure}
  \begin{center}
  \FigureFile(80mm,80mm){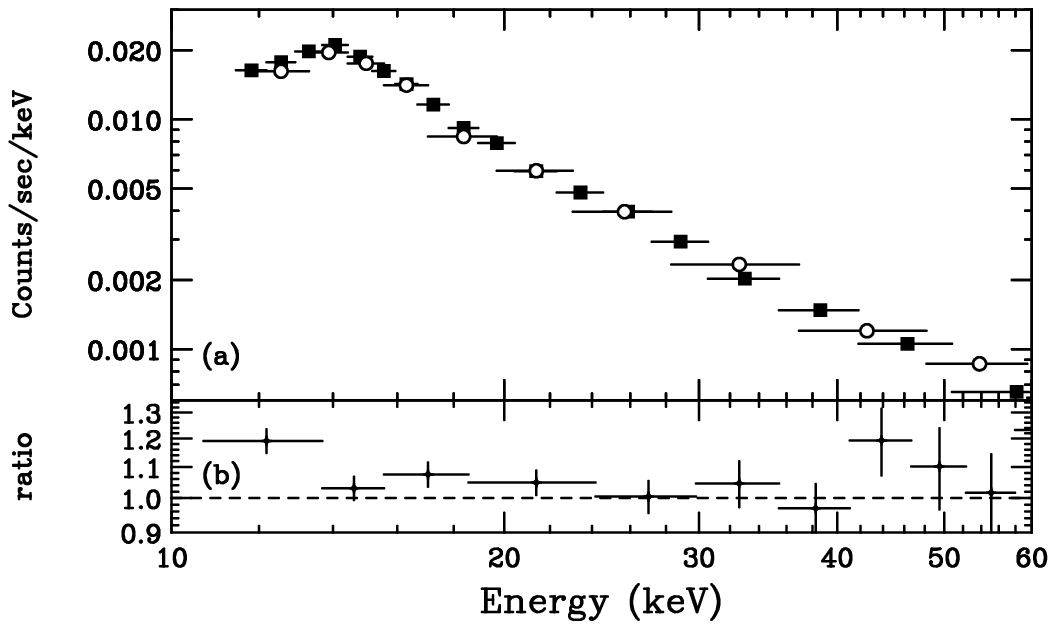}\\
  \end{center}
  \caption 
{(a) Raw HXD-PIN spectra (open circles) observed during the Earth-occultation periods, 
compared with the NXB model predictions (filled squares). (b) The model to data ratio.}
\label{n4258-earth}
\end{figure}

To further evaluate the systematic error of the NXB, we accumulated HXD-PIN spectrum 
while the satellite was pointing to the earth, with a exposure of 9.3 ks, and compared the results in figure 2 with the NXB model prediction. 
Thus, the modeled NXB spectrum is slightly higher than the Earth spectrum, 
especially around 12--20 keV. 
A more quantitative result of this comparison is given in table 1, where we find that the 
NXB is overestimated by $\sim$ 3\% in the 12--40 keV range, even though this is within the guaranteed 
accuracy of the present model (Fukazawa et al. 2008).  

\begin{table*}
 \caption{Signal and background count rates from HXD-PIN during the on-source exposure.}
 \label{tbl4258}
 \begin{center}
  \begin{tabular}{lrrr}
   \hline\hline
   Component & 12-25 keV & 25-40 keV & 40-60 keV \\
   \hline
   On-source      & 0.3944 $\pm$ 0.0021  
                       & 0.0925 $\pm$ 0.0010   
                       & 0.0463 $\pm$ 0.0007 \\
                       
   NXB$^*$        & 0.3445 $\pm$ 0.0048
                       & 0.0832 $\pm$ 0.0004
                       & 0.0425 $\pm$ 0.0012 \\

   CXB$^\dagger$              & 0.0225 
                       & 0.0040 
                       & 0.0005\\[1.5ex]
                       
  Net Signal$^\ddagger$        & 0.0274 $\pm$ 0.0022
                       & 0.0053 $\pm$ 0.0010
                       & 0.0035 $\pm$ 0.0007\\
                       
   & $\pm$ 0.0048 
                       & $\pm$ 0.0004
                       & $\pm$ 0.0012\\

      \hline\hline
\end{tabular}
\end{center}
\begin{itemize}
\item[$^*$] Modeled with LCFITDT (Fukazawa et al. 2008). The systematic errors are described in section 2.4.  
\item[$^\dagger$]  Predicted from the HEAO-1 measurement (Gruber et al. 1999).  
\item[$^\ddagger$]  Net count rate obtained by subtracting the NXB and CXB from the on-source count rate. 
The first and the second uncertainties represent the statistical and systematic errors, respectively. 
\end{itemize}
\end{table*}

The counts remaining after subtracting the NXB still include the contribution of the cosmic X-ray background (CXB; \cite{Boldt1987}), 
which must be subtracted as well. 
The CXB contribution was estimated using the HXD-PIN response to diffuse sources, assuming the spectral CXB surface brightness model determined by 
HEAO 1 (\cite{Gruber1999}): 
9.0$\times$10$^{-9}$ ($E$/3 keV)$^{-0.29}$
 $\exp$ (-$E$/40 keV) erg 
cm$^{-2}$ s$^{-1}$ str 
$^{-1}$ keV$^{-1}$,
where $E$ is the photon energy. 
The estimated CXB count rate is 5\%
of the NXB signals. 
The actual CXB
normalization has recently been found (\cite{Revnivtsev2003}; \cite{Churazov2007}) to be higher by 10--15\% at 10--100 keV than
that of Gruber et al. (1999), 
but the difference is negligible because it is within the systematic uncertainty of the NXB subtraction. 
As indicated by table 2, 
the source has been detected significantly at least up to $\sim$ 40 keV with a significance level of $>$ 5 $\sigma$.

In contrast to the detections with HXD-PIN, 
the source was not detected with HXD-GSO. 
The estimated upper limit, $\sim$ 1m Crab at $\sim$ 50 keV, is consistent with an extrapolation from the HXD-PIN signals 
which indicate a source intensity of $\sim$ 0.6 mCrab at 30 keV.

\subsection{Light Curves
\label{subsec:obs_ltcurve}}
\begin{figure}
  \begin{center}
    \FigureFile(80mm,80mm)
   {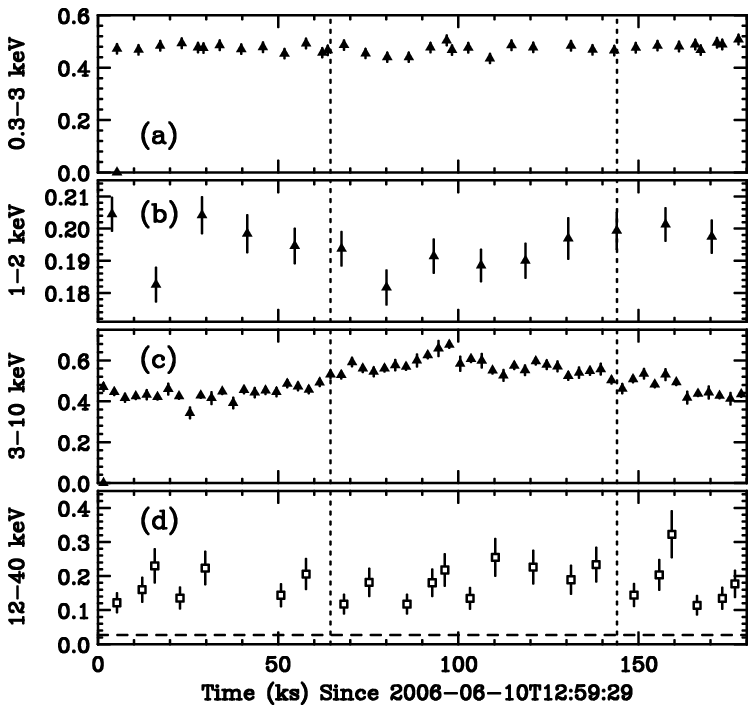}\\
  \end{center}
  \caption
 {Background-subtracted light curves of NGC~4258
 taken with Suzaku. 
Panels (a), (b), (c), and (d) refer to those obtained with XIS-FI (0.3--3.0 keV), XIS-FI (1.0--2.0 keV), 
XIS-FI (3.0--10 keV), and HXD-PIN (12--40 keV), respectively. 
The bin widths in panels (a) through (d) are 2.5 ks, 10 ks, 3 ks, and 2.5 ks, respectively. 
The HXD-PIN result is presented after subtracting the NXB as described in section 2.2, and correcting for dead times. 
It includes the CXB contribution by 0.02 c s$^{-1}$ (a dashed line in panel d). 
In contrast, the XIS light curves include NXB. 
}\label{n4258-lc}
\end{figure}

Figure \ref{n4258-lc} shows XIS-FI and HXD-PIN light
curves of NGC~4258 from the present observation. 
In the 3--10 keV band, the source is gradually increasing and then decreasing by
$\pm$10\% on time scales of $\sim$ 100 ks. 
In contrast, the light curves below 3 keV is much closer to being constant.
These agree with the previous reports (e.g., Makishima et al. 1994) that the dominant emission in this energy range is thermal diffuse emission. 
Interestingly, some variations are suggested in the 1--2 keV band. 
Although the HXD-PIN light curve has rather poor photon statistics, 
it is inconsistent with being constant, 
because fitting it with a constant yields $\chi^2 \sim$ 79 for 24 d.o.f. 
 
\subsection{Time-Averaged Spectra
\label{subsec:obs_avespec}} 

\begin{figure}
  \begin{center}
    \FigureFile(80mm,80mm)
{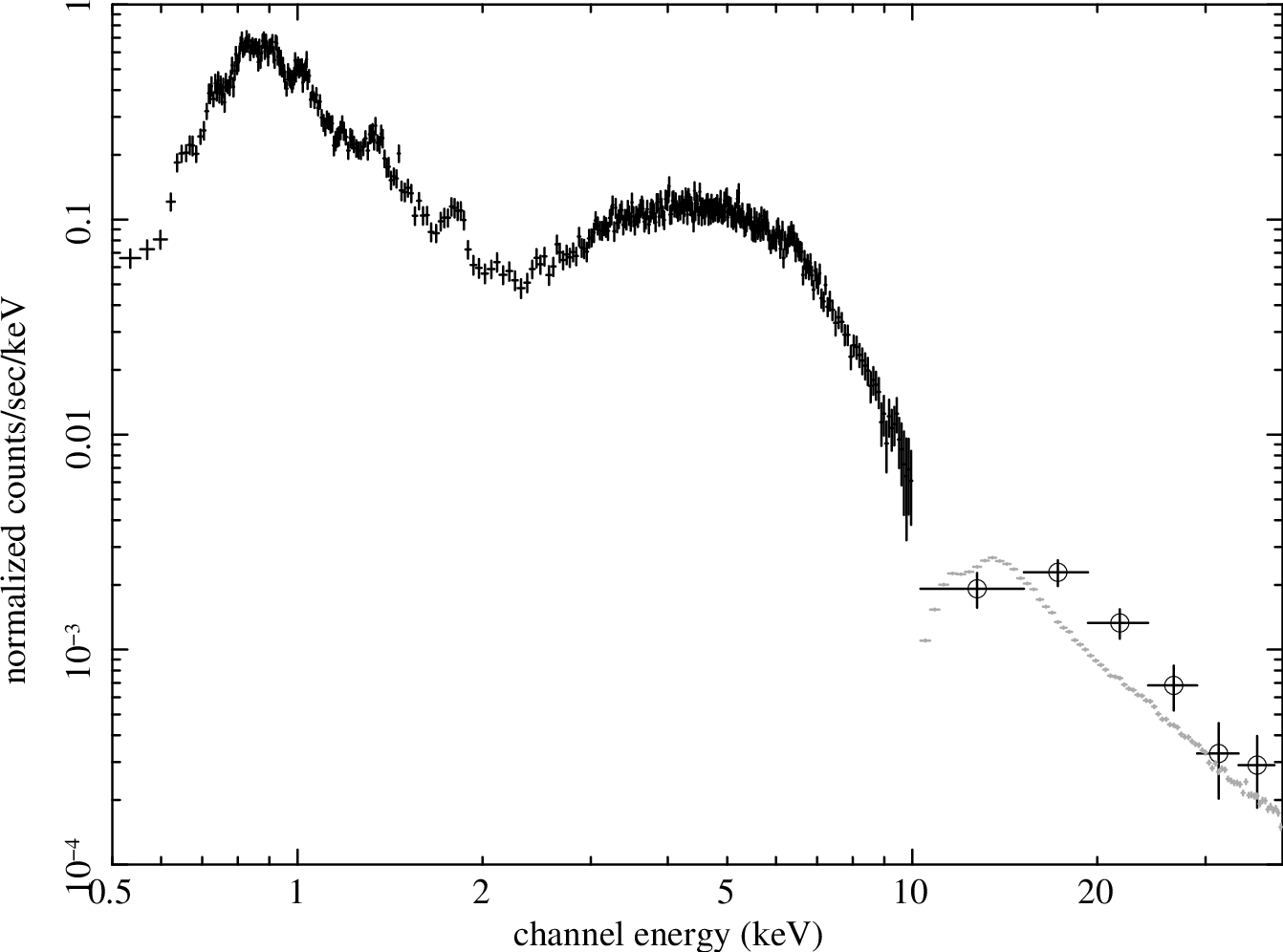}\\
  \end{center}
  \caption
 {Time-averaged XIS-FI (cross) and HXD-PIN (circle) spectra of the nuclear region of NGC~4258. 
The XIS spectrum is presented after subtracting
the NXB, and the HXD-PIN data after subtracting
the NXB + CXB. 
Both spectra are shown without removing the instrumental responses.
For comparison, the 5\% level of the modeled NXB of HXD-PIN
is indicated in gray.}
\label{obsspec}
\end{figure}

\begin{figure}
  \begin{center}
        \FigureFile(80mm,80mm)
   {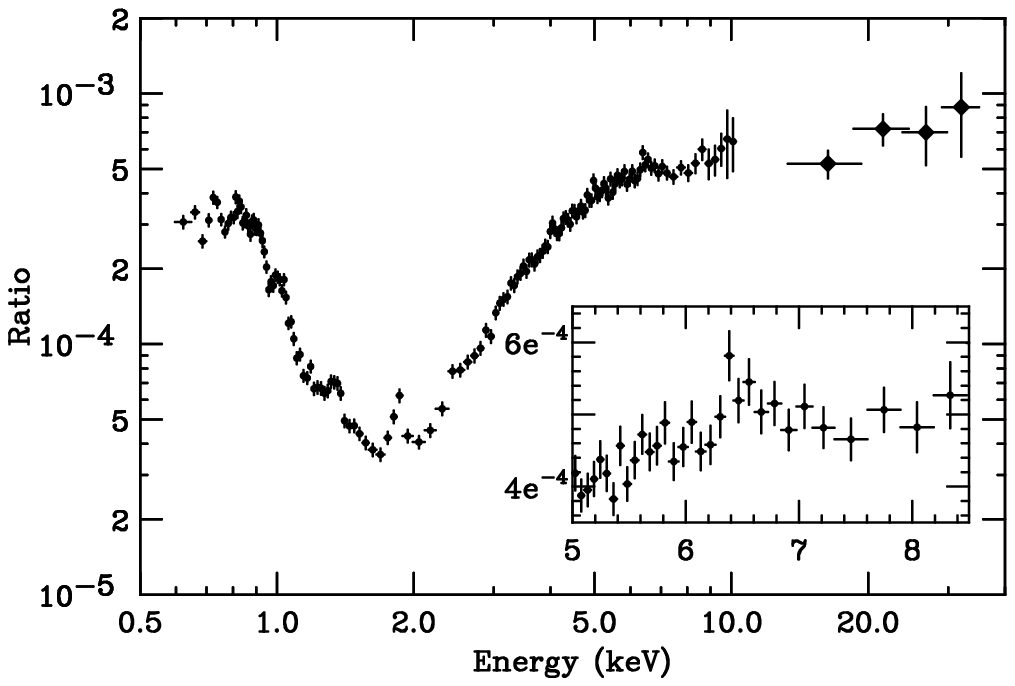}\\
  \end{center}
  \caption
 {The same spectrum as figure 3, normalized to those of the Crab Nebula acquired on 2005 September 15. The 
 inset shows details around the Fe-K lines and edges.}
\label{n4258-crab}
\end{figure}

Figure \ref{obsspec} shows time-averaged Suzaku spectra of
NGC~4258. The HXD-PIN spectrum is corrected for the dead time, and is presented after 
subtracting the backgrounds (NXB + CXB) as described in section 2.2. 
As revealed with ASCA (Makishima et al. 1994) and BeppoSAX (Fiore et al. 2001), 
the XIS spectrum thus consists of at least two components, a soft thermal and the absorbed hard ones, dominant in energies 
below and above $\sim$ 2 keV, respectively. 
To grasp overall features of the spectra, 
we divided them by those of the Crab Nebula, 
and show the results in figure \ref{n4258-crab}. 
The ratio clearly reveals the two components mentioned above.
However, the flux decrease from $\sim$ 4 keV to $\sim$ 2 keV is less steep than is described with a single-valued photoelectric absorption.  
It suggests that another component of a medium hardness (\cite{Max1994}) extends up to $\sim $3 keV. 
The Fe-K emission line is rather weak, and Fe-edge 
is not prominent. 

\begin{figure}
  \begin{center}
          \FigureFile(80mm,80mm)
   {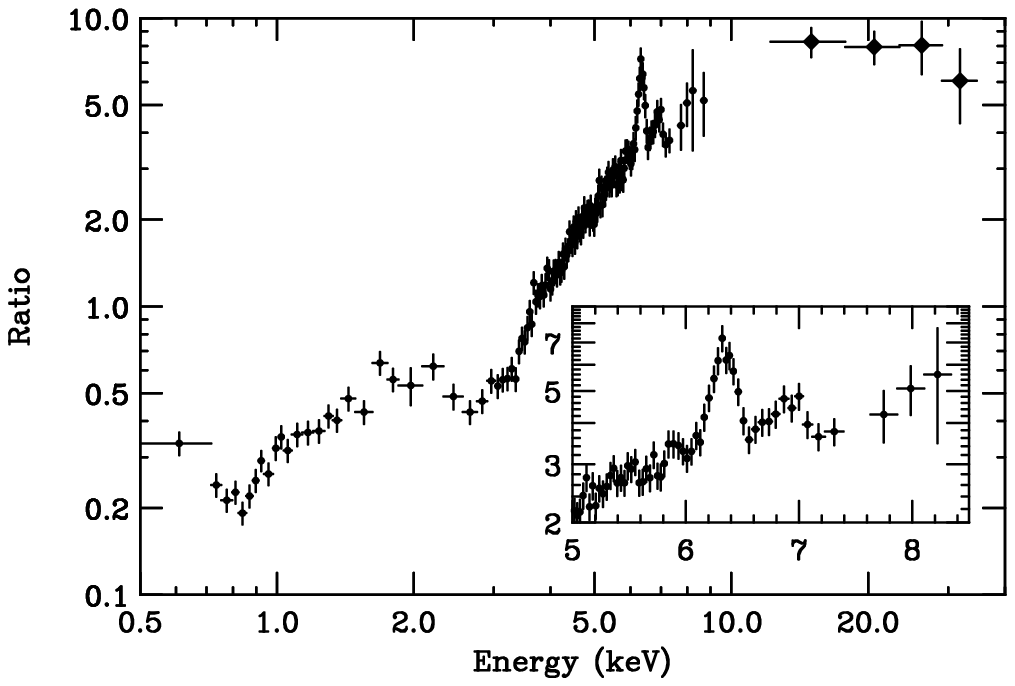}\\
  \end{center}
  \caption
 {Suzaku spectra of NGC~4388 (Shirai et al. 2008), divided by those of NGC~4258 presented in figure \ref{obsspec}. 
 The inset shows details around the Fe-K lines and edges.}
\label{n4258-4388}
\end{figure}

Before conducting spectral fitting analysis, let us compare the spectra of NGC~4258 
with those of NGC~4388, one of the typical Compton-thick AGNs.   
Specifically, we divided the Suzaku spectra of NGC~4388 (\cite{Shirai2008}) by those of NGC~4258, 
and show the ratios in figure \ref{n4258-4388}. 
The result reveals an intense Fe-K line and a prominent Fe-K edge, 
implying that these features are much stronger in NGC~4388 than in NGC~4258. 
Although the ratio increase from $\sim$ 3 keV to $\sim$ 10 keV is primarily understood as due to 
a heavier absorption in NGC~4388, 
the energy dependence is approximately power-law like. 
The statement would be reinforced if the possible contribution from a medium-hardness component in NGC~4258 is considered.  
This is presumably due to the strong reflection component in NGC~4388: 
conversely, we infer that the corresponding component is significantly weaker in NGC~4258. 
This inference is consistent with a slight decrease of the ratio from 20 keV to 30 keV, 
to be ascribed to the strong Compton reflection hump in NGC~4388 and its weakness in NGC~4258.

\section{Spectral Analysis}
\label{sec:ana_avspec}

\subsection{{Analysis of the Time-Averaged 0.9--40 keV Spectra}
\label{subsec:aba_ave}}
To quantify the spectrum, 
we assign the PIN data points with a systematic error of 20\%,
which corresponds to an NXB uncertainty of $\sim$ 1.4\%.
At each spectral bin,
this systematic error is added in quadrature to the statistical error.
In the following spectral analysisde
we constrain each spectral model parameter
to be the same among XIS-FI, XIS-BI, and HXD-PIN.
The overall model normalization is set identical between XIS-FI and XIS-BI,
while 13 \% higher for HXD-PIN (Suzaku memo 2008-06).
Furthermore, we ignore the 1.84--1.86 keV range
due to the XIS response uncertainty around the Si edge.

Although our main interest is in the absorbed nuclear emission,
its low-energy end ($\lesssim 4$ keV) is confused with other softer spectral components.
To disentangle this effect,
we start from a relatively wide energy range of 0.9--40 keV,
discarding the softer ranges to avoid the complexity due to very soft components.
Employing the procedure described above,
we hence fitted simultaneously the XIS (FI and BI) and HXD-PIN spectra,
in the 0.9--9.0 keV, 0.9--8.0 keV, and 12.0--40 keV energy ranges, respectively.
Figure \ref{widebandspec}a shows the case when we employed the simplest fitting model
that consists of an absorbed power-law component
and a single-temperature plasma emission model,
representing the absorbed nuclear emission and the extended soft thermal emission, respectively.
The plasma emission was modeled using \texttt{apec} code
with a free temperature and a free overall metallicity
(but keeping the solar abundance ratios),
and was subjected to the Galactic absorption column of $1.2 \times 10^{20}$ cm$^{-2}$.
The fit then gave, as the best estimates, a power-law photon index of $\Gamma \sim 1.93$,
its absorbing column of $N_{\rm H} \sim 1.1 \times 10^{23}$ cm$^{-2}$,
and a plasma temperature and abundance of 0.58 keV and 0.47 solar, respectively.
However, as evident in figure \ref{widebandspec}a, the fit was not acceptable,
leaving significant residuals in the energy range of 1.5--2.5 keV
where the two model components cross over.
In addition, the fit residuals reveal a weak Fe-K line at $\sim 6.4$ keV.

As already mentioned in section 2.5 referring to figure \ref{n4258-crab},
the 1.5--2.5 keV residuals indicate the presence of
another spectral component with an intermediate hardness;
\citet{Max1994} already noticed the same component,
and attributed it to the integrated emission from point sources.
Our results support this interpretation,
because the very strong He-like Si K$_\alpha$ line (at 1.86 keV)
and  the weakness of its H-like counterpart (at 2.01 keV)
imply that any hotter thin-thermal emission is unlikely
to contribute significantly to our spectrum
in energies above $\sim 2$ keV.

\begin{table*}
 \caption{The best-fit parameters to the time-averaged Suzaku spectra, with 90 \%-confidence error ranges. }
 \label{all_tbl}
 \begin{center}
  \begin{tabular}{llrrrr}
   \hline\hline
   Component & Parameter & \multicolumn{2}{c}{0.9--40 keV} & \multicolumn{2}{c}{1.7--40 keV}    \\
                                   &     &  &                  & NXB nominal & -2\% NXB \\

   \hline
   PL     & $\Gamma$%
                       & $1.93^{+0.05}_{-0.04}$ 
					   & $1.93 \pm 0.01$
                       & $1.88 \pm 0.01$
                       & $1.86^{+0.01}_{-0.02}$  \\

                    & $N_\mathrm{PL}$$^*$%
                       & $4.62^{+0.01}_{-0.02}$
                       & $4.75^{+0.67}_{-0.05}$ 
                       & $4.21^{+0.01}_{-0.30}$
                       & $4.22^{+0.01}_{-0.40}$ \\

                    & $N_\mathrm{H}$$^\dagger$%
                       & $11.1^{+0.3}_{-0.8}$
	                   & $11.2 \pm 0.2$
                       & $10.7^{+0.3}_{-0.2}$
                       & $10.7^{+0.2}_{-0.2}$   \\[1.5ex]

   Fe~I~K$\alpha$   & $E_\mathrm{c}$$^\ddagger$ %
                       & $6.39^{+0.04}_{-0.03}$
                       & $6.39^{+0.04}_{-0.03}$
                       & $6.39^{+0.04}_{-0.03}$
                       & $6.39^{+0.05}_{-0.04}$ \\

                    & $I$ $^\S$ %
                       & $6.8^{+0.9}_{-1.4}$
                       & $6.8^{+0.9}_{-1.9}$
                       & $6.5^{+0.8}_{-2.4}$
                       & $6.8^{+0.7}_{-2.5}$       \\

                    & EW (eV) 
                      & $53.3^{+7.1}_{-11.0}$
                      & $51.2^{+7.1}_{-14.3}$
                      & $51.4^{+6.3}_{-19.0}$
                      & $51.0^{+5.2}_{-18.8}$ \\[1.5ex]

   bremss     & $kT_{\rm{br}}$ (keV) %
                       & $11.1^{+19.1}_{-6.2}$
                       & 5 (fix)
                       & 5 (fix)
                       & 5 (fix) \\
                    & $L$ $^\|$%
                       & $4.9 \pm 0.1 $  
                       &  $4.9 \pm 0.1 $
                        & $4.9^{+0.7}_{-1.5}$
                        & $4.8^{+0.9}_{-1.4}$   \\[1.5ex]
   apec     & $kT$ (keV) %
                       & $0.58 \pm 0.01$
                       & $0.58 \pm 0.01$
                       & $0.78^{+0.04}_{-0.05}$
                       & $0.76^{+0.04}_{-0.05}$ \\
                    & Abundance%
                       & $0.47^{+0.17}_{-0.13}$
                       & 1 (fix)
                       & 1 (fix)
                       & 1 (fix)  \\
                    & $L$ $^\|$%
                       & $4.3^{+24.5}_{-1.72}$
                       & $0.22 \pm 0.03$
                       & $0.36 \pm 0.03$
                       & $0.39 \pm 0.03$  \\[1.5ex]
   Reflection       & $f_\mathrm{refl}$%
					&
                     &
                    & $< 0.3$
                       & $< 0.3$ \\[1.5ex] \\

   $\chi^{2}$/d.o.f. &   & 734.7/619  & 738.3/618 & 512.0/449  & 494.6/449   \\
      \hline\hline
\multicolumn{5}{@{}l@{}}{\hbox to 0pt{\parbox{180mm}{\small
\begin{itemize}
\item[$^*$] The power-law normalization at 1 keV, in units of $\mathrm{10^{-3}~photons~keV^{-1}~cm^{-2}~s^{-1}}$~at 1 keV.
\item[$^\dagger$] Equivalent hydrogen column density in  $10^{22}$ cm$^{-2}$.
\item[$^\ddagger$] Center energy in keV.
\item[$^\S$] Photon number flux in $10^{-6}$~photons~cm$^{-2}$~s$^{-1}$.
\item[$^\|$] Integrated 2--20 keV luminosity at 7.2 Mpc in units of $10^{39}$ erg s$^{-1}$.
\end{itemize}
     }\hss}}
  \end{tabular}
 \end{center}
\end{table*}

To improve the fit,
we added a thermal bremsstrahlung model and a single Gaussian.
The former is an empirical modeling of point sources
(mostly thought to be low-mass X-ray binaries) after \citet{Max1994};
we leave its temperature and normalization both free,
and fix its absorption artificially at the value of
 $2 \times 10^{21}$ cm$^{-2}$ \citep{Max1989}
for a better approximation of integrated point-sources.
The latter is allowed to have a free centroid and free intensity,
but is assumed to be narrow ($\sigma = 0.001$ eV).
Then, as shown in figure \ref{widebandspec}b,
the fit was significantly improved.
As summarized in table~\ref{all_tbl},
the temperature of the thermal bremsstrahlung was obtained
 with large errors as $kT_{\rm{br}} = 11.1^{+19.1}_{-6.2}$ keV.
This is because $kT_{\rm{br}}$ correlates negatively
with the abundance of the thermal component,
in such a way that a lower value of $kT_{\rm{br}}$ enhances its soft X-ray contribution,
and hence reduces the continuum attributable to the plasma emission.
If, e.g., the plasma abundance of the soft thermal component is fixed at 1.0 solar,
we obtain $kT_{\rm{br}}$ = $5.2^{+2.9}_{-3.0}$ keV,
which is reasonable as a summed point-source spectrum
from a spiral galaxy {\citep{Max1989}.
In addition, the 2.0--20 keV luminosity of this component,
 $5.8^{+2.9}_{-3.3} \times 10^{39}$ erg s$^{-1}$,
is consistent with that interpretation \citep{Max1989}.
We therefore choose the 1.0 solar abundance and $kT_{\rm{br}} = 5.0$ keV
as our baseline modeling.
The model parameters obtained under these conditions
are also given in table~\ref{all_tbl}.

\subsection{{Time-Averaged Spectra above 1.7 keV}
\label{subsec:above1.7}}
Although the above 4-component model is approximately successful,
the fit is not yet fully acceptable;
this would affect our error estimate.
Since the fit residuals are seen in soft energies below $\sim$ 2 keV,
they are likely to be attributed to such effects as deviations of the plasma emission
from the assumed isothermally and solar-ratio composition.
Instead of trying to solve these issues, we decided to limit the fitting range to above 1.7 keV.
The plasma abundance was again fixed at 1 solar, together with $kT_{\rm{br}}$ = 5.0 keV.
Then, the fit was significantly improved to $\chi^2/\nu = 518.4/449$.
As summarized in table~\ref{all_tbl},
the fit became still better,  $\chi^2/\nu = 512.0/449$,
by assigning a systematic error of 1\% (Suzaku memo 2008-06) to the model.

As shown in figure \ref{widebandspec}c,
we still observe a negative deviation around 12.0--25 keV in the HXD-PIN data,
which could be due to the overestimated NXB in this energy range
as noted in section \ref{subsec:obs_hxd}
based on our Earth-occultation calibration.
We therefore changed the NXB intensity to be subtracted
by several percent from the nominal value,
and repeated the fitting.
Then, just as expected,
the case of 2\% reduced NXB made the fit acceptable with $\chi^2/\nu = 494.6/449$.
As presented in figure 7d, the deviation around 12--40 keV apparently decreased.
We therefore adopt this case as our best modeling,
and summarize the parameters in table \ref{all_tbl}.
Figure~\ref{ape_eeuf} shows the same spectra
in the deconvolved $\nu F \nu$ form,
where the deconvolution utilizes  the  best-fit model.

Under the above best-fit model,
the nuclear emission has been modeled by an absorbed power law
with a photon index of $\Gamma \sim 1.86$,
absorbed by a column density of $N_\mathrm{H} \sim 1.07 \times 10^{23}$~cm$^{-2}$.
The temperature of the thermal plasma was determined as $\sim$ 0.76 keV,
mainly by the intensity ratio between the He-like Si and H-like Si lines.
The obtained unabsorbed 2.0--10.0 keV flux of the nuclear
power-law component  is $1.35 \times 10^{-11}$~erg~cm$^{-2}$~s$^{-1}$,
corresponding to a luminosity of $7.9 \times 10^{40}$~erg~s$^{-1}$ at 7.2~Mpc.
The narrow Gaussian emission line was obtained
at $\sim 6.4$~keV  with an equivalent width of $\sim$ 50 eV.
In order to estimate the line width,
we allowed the Gaussian to be broad,
but this did not improve the fit significantly.

Although the data have already been reproduced successfully with this simple model,
an AGN spectrum generally bears a feature
due to ``reflection" from circum-nuclear cold materials.
To evaluate its possible contribution to the NGC~4258 spectrum,
we tentatively added a reflected continuum component (\texttt{pexrav} in XSPEC)
to our fitting model.
The input to \texttt{pexrav} was assumed to be a power-law 
with the same photon index as the intrinsic power-law, without a cutoff.
The reflector was assumed to have solar abundances,
and an inclination of $i=60^{\circ}$ as a representative case.
The inclusion of this additional component did not improve the fit,
and its strength was constrained as $\ensuremath{f_\mathrm{refl}}<  0.3$ at the
90 \% confidence limit (the best-fit being at $\ensuremath{f_\mathrm{refl}} = 0.063$ with $\chi^2/\nu = 494.5/448$).
In order to understand how the data constrain $\ensuremath{f_\mathrm{refl}}$,
we repeated the fitting with $\ensuremath{f_\mathrm{refl}}$ purposely fixed at 1.0,
simulating a reflector with an infinite slab geometry.
Then, as shown in figure \ref{widebandspec}e,
the HXD-PIN data became rather discrepant with the model, yielding $\chi^2/\nu = 527.4/449$.
In other words, HXD-PIN would have to be measuring higher fluxes
if the nuclear emission were accompanied
by a reflection component with $\ensuremath{f_\mathrm{refl}} \sim 1$.
Thus, the HXD data play an essential role in constraining the reflection.
To confirm that these results are not affected by the uncertainty in $kT_{\rm{br}}$,
we changed it to 2.0 keV and 10.0 keV, but both yielded $\ensuremath{f_\mathrm{refl}} < 0.3$.
For a further confirmation, we used the nominal NXB model
instead of the 2\% reduced one,
but the upper limit on $\ensuremath{f_\mathrm{refl}}$ remained unchanged.

The above choice of $i=60^\circ$ is considered appropriate for a general case,
where reflecting materials assume a shape of an inflated torus
as dictated by the Unified Scheme.
However, the strong constraint on $\ensuremath{f_\mathrm{refl}}$ derived under this assumption suggests 
that the material is in reality localized in a limited solid angle containing our line of sight,
like in the case of NGC~4945 \citep{Itoh2008}.
To examine the consistency of this alternative configuration,
we repeated the fitting assuming the reflector to have $i=83^\circ$,
which is the same as the inclination of the water-maser emitter \citep{Miyoshi1995}.
Then, the constraint has been relaxed to $\ensuremath{f_\mathrm{refl}}<2.0$,
and the case of $\ensuremath{f_\mathrm{refl}}=1.0$ (a flat infinite slab) became acceptable with $\chi^2 = 498.7$.
In other words, the data are consistent with the presence
of a slab-like reflector viewed nearly sideways.
This result is reasonable,
because signals from such a flat reflector
will scale as $\sim \cos i$ in the observed spectra.
Finally, we let $i$ and $\ensuremath{f_\mathrm{refl}}$ both float,
and found that the data favor $i \sim 30^{\circ}$
but the fit does not improve ($\chi^2/\nu = 494.5/447$).

\begin{figure*}
	\begin{center}
    \FigureFile(160mm,160mm){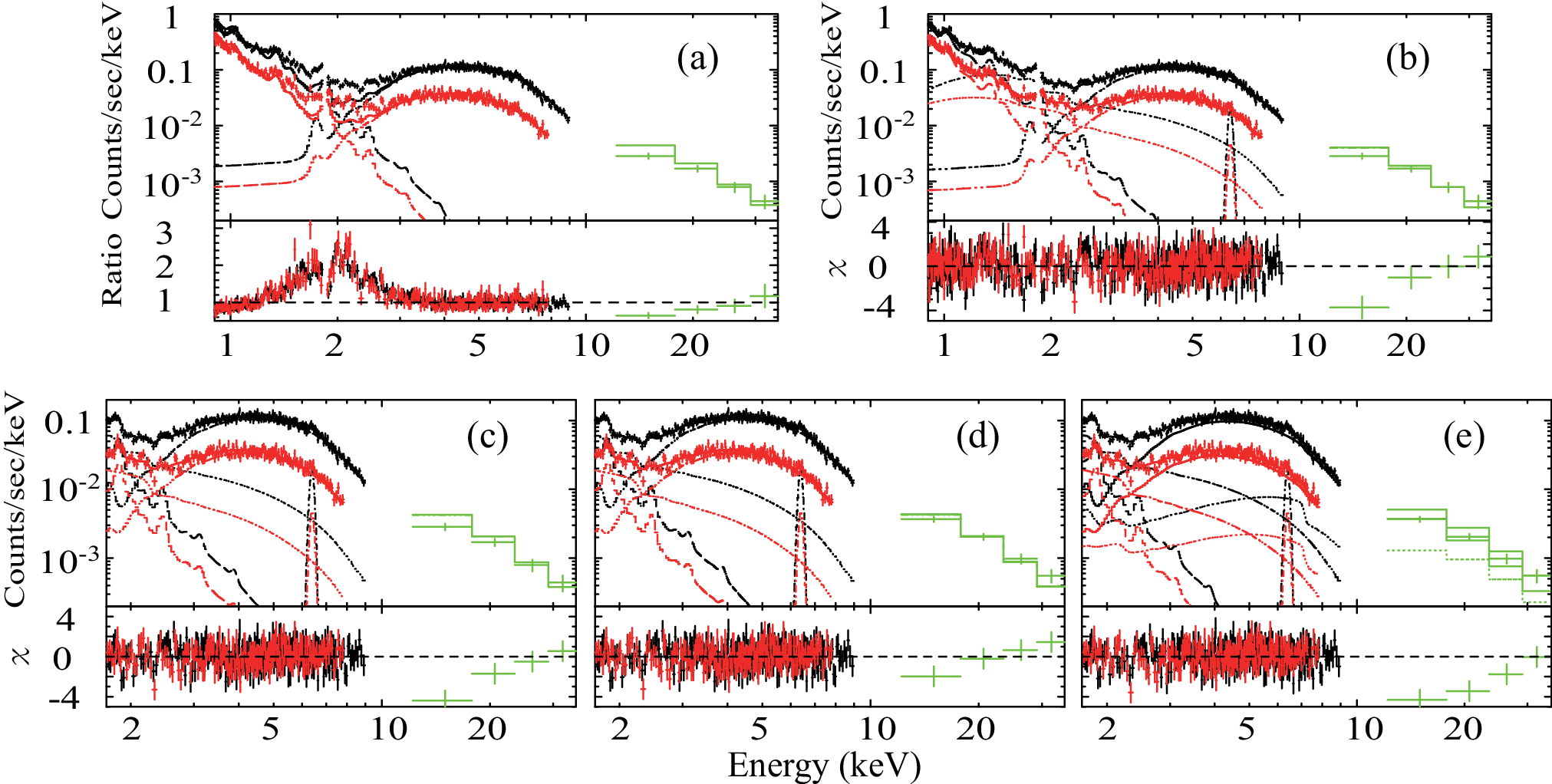}
\end{center}
  \caption{Wide-band modeling of the background-subtracted XIS
  (black for FI and red for BI) and  HXD-PI (green) spectra of NGC~4258.
 The employed energy range is 0.9--40.0 keV in panels (a) and (b),
 while 1.7--40 keV in panels (c), (d), and (e).
(a) A fit with an absorbed power-law and a single temperature \texttt{apec} model.
(b) An improved fit obtained by adding a bremsstrahlung component and a Gaussian.
(c) A fit with the same model as panel (b), but using  a narrower energy rage. 
(d) The NXB level to be subtracted is reduced by 2\%.
(e) The same as panel (d), but a reflection component with $\ensuremath{f_\mathrm{refl}} = 1.$0 is added.}
\label{widebandspec}
\end{figure*}

\begin{figure}
  \begin{center}
  \FigureFile(80mm,80mm)
{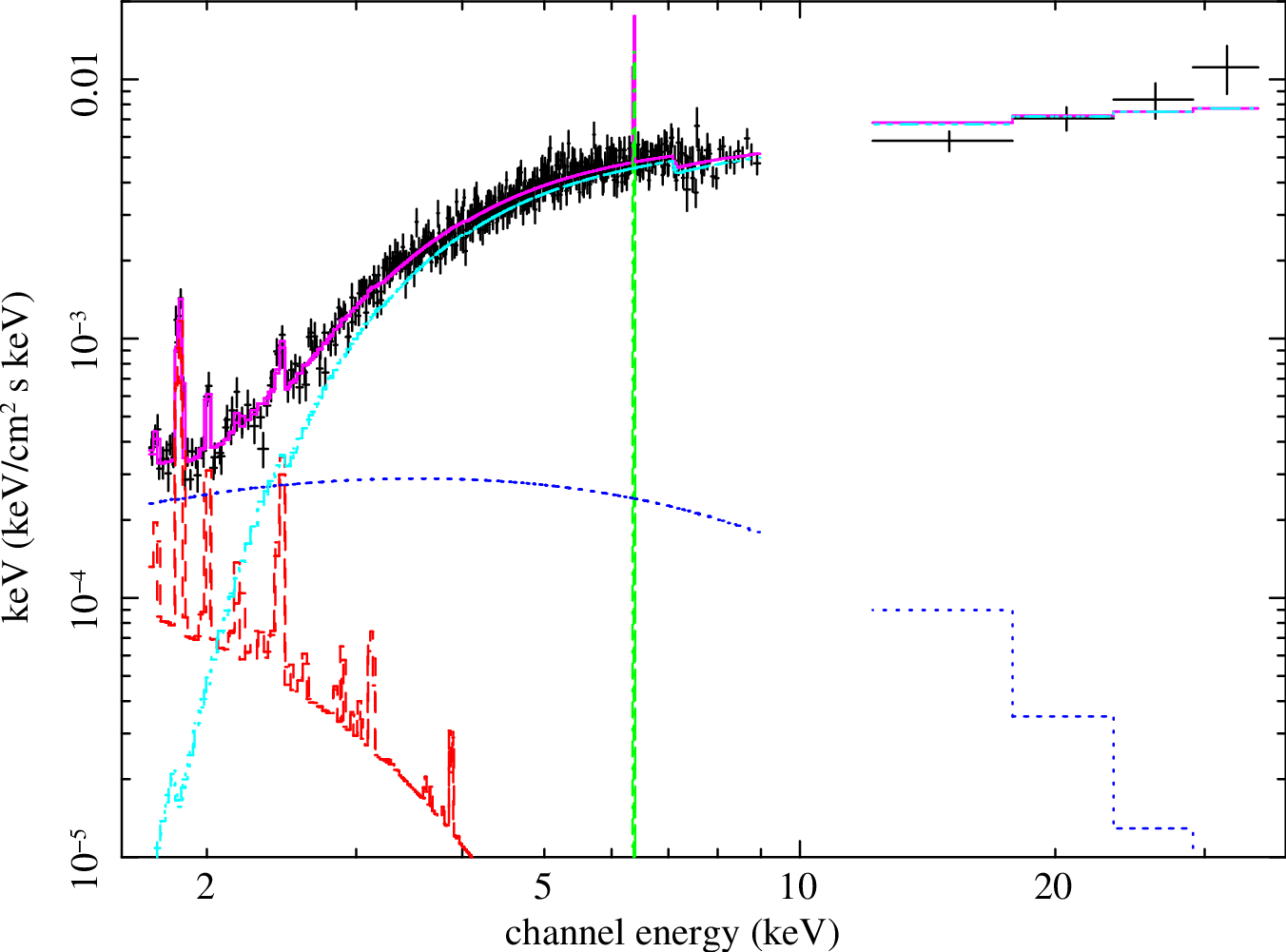}\\
  \end{center}
  \caption{The same spectra of NGC~4258  as in figure 7d,
 but presented in the deconvolved $\nu F \nu$ form.
The observed data are shown in black, in comparison with the best-fit model in magenta.
The thermal plasma, bremsstahlung, narrow iron line, and the absorbed power-law
components are shown in red, blue, green, and cyan, respectively}
\label{ape_eeuf}
\end{figure}

\subsection{{Analysis of Time Variations}
\label{subsec:hl_ave}}
During the present observation,
the absorbed nuclear X-rays exhibited mild intensity variations (figure \ref{n4258-lc}).
In order to search the spectra for intensity-correlated changes,
we divided the entire observation period
into two subsets, namely ``high-flux'' and ``low-flux'' phases,
when the 2---10 keV count rate is higher
and lower than 0.5 cts~s$^{-1}$, respectively.
The two dotted lines in figure \ref{n4258-lc} separate them;
the high-flux phase between the lines and the low-flux phase outside.
The net exposure is
42.9~ks (XIS) and 38.7~ks (HXD) for the high-flux phase,
and 57.0~ks (XIS) and 52.9~ks (HXD) for the low-flux phase.

\begin{figure}
  \begin{center}
         \FigureFile(80mm,80mm) 
{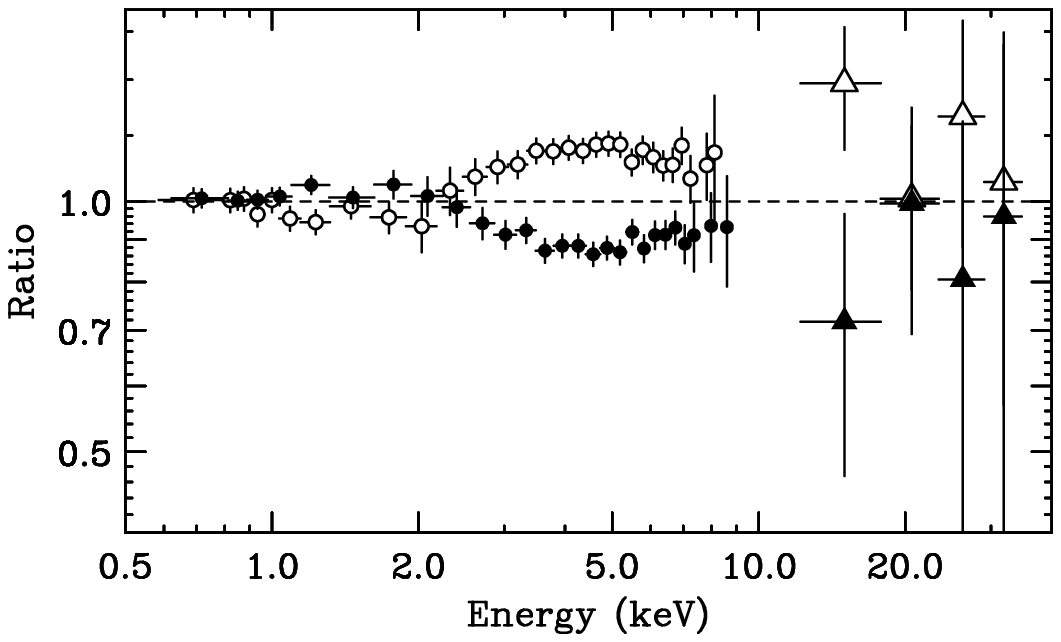}\\
  \end{center}
  \caption{Spectra from the ``high-flux'' (open) and ``low-flux'' (filled) phases,
  shown as their ratios to the time averaged spectrum.
  The XIS and PIN data points are shown in circles and triangles, respectively.
  }\label{n4258-hilo_ratio}
\end{figure}

Figure~\ref{n4258-hilo_ratio} shows the spectra from these two phases,
divided by the time averaged spectrum.
Thus, the variation is evident in energies above $\sim$ 2.5 keV,
in agreement with a simple idea
that only the absorbed nucleus should vary.
Interestingly, however, we observe small but significant variations
in the 1.0---2.0 keV range, apparently in anti-phase with the dominant hard X-ray variations.
This is also suggested by figure 3.

\begin{table*}
 \caption{The best-fit parameters to the high-flux and low-flux phase spectra over 1.7--40 keV of NGC~4258.*}
 \label{hl_tbl}
 \begin{center}
 \begin{tabular}{lcccc}
   \hline\hline
   Component & Parameter & High & Low & difference\\
   \hline
   Intrinsic PL     & $\Gamma$%
                       & $1.96^{+0.03}_{-0.02}$
                       & $1.70 \pm 0.02 $
		         & $2.8^{+0.74}_{-0.61}$ \\

                    & $N_\mathrm{PL}$ %
                       & $5.95^{+0.98}_{-0.07}$
                       & $2.73 \pm 0.04$
                       & $7.5^{+23.5}$  \\
                    & $N_\mathrm{H}$ %
                       & $11.3 \pm 0.3$
                       & $9.4 \pm 0.3$
                       & $18.1^{+5.3}_{-4.2}$    \\[1.5ex]

   Fe~I~K$\alpha$   & $E_\mathrm{c}$%
                       & \multicolumn{2}{c}{$6.39 \pm 0.05$} \\

                    & $I$
                       & \multicolumn{2}{c}{$5.93^{+1.45}_{-1.35}$} \\

   bremss     & $kT$ (keV) %
                       & \multicolumn{2}{c}{5.0 (fix)}  \\

                    & $L$ %
                       & \multicolumn{2}{c}{$4.3^{+0.3}_{-0.25}$} \\

   apec     & $kT$ (keV) %
                       &  \multicolumn{2}{c}{0.76 (fix)} \\
                    & Abundance%
                       & \multicolumn{2}{c}{1.0 (fix)}    \\
                    & $L$ %
                       & \multicolumn{2}{c}{0.39 (fix)}  \\

   $\chi^{2}$/d.o.f. &   & \multicolumn{2}{c}{435.8/436}  & 45.2/42\\
      \hline\hline
\multicolumn{4}{@{}l@{}}{\hbox to 0pt{\parbox{180mm}{\small
\par\noindent\\
*All physical quantities and their units are the same as in table 3.
  \par\noindent
\par\noindent
     }\hss}}
  \end{tabular}
 \end{center}
\end{table*}

\begin{figure}
  \begin{center}
   \FigureFile(80mm,80mm) 
{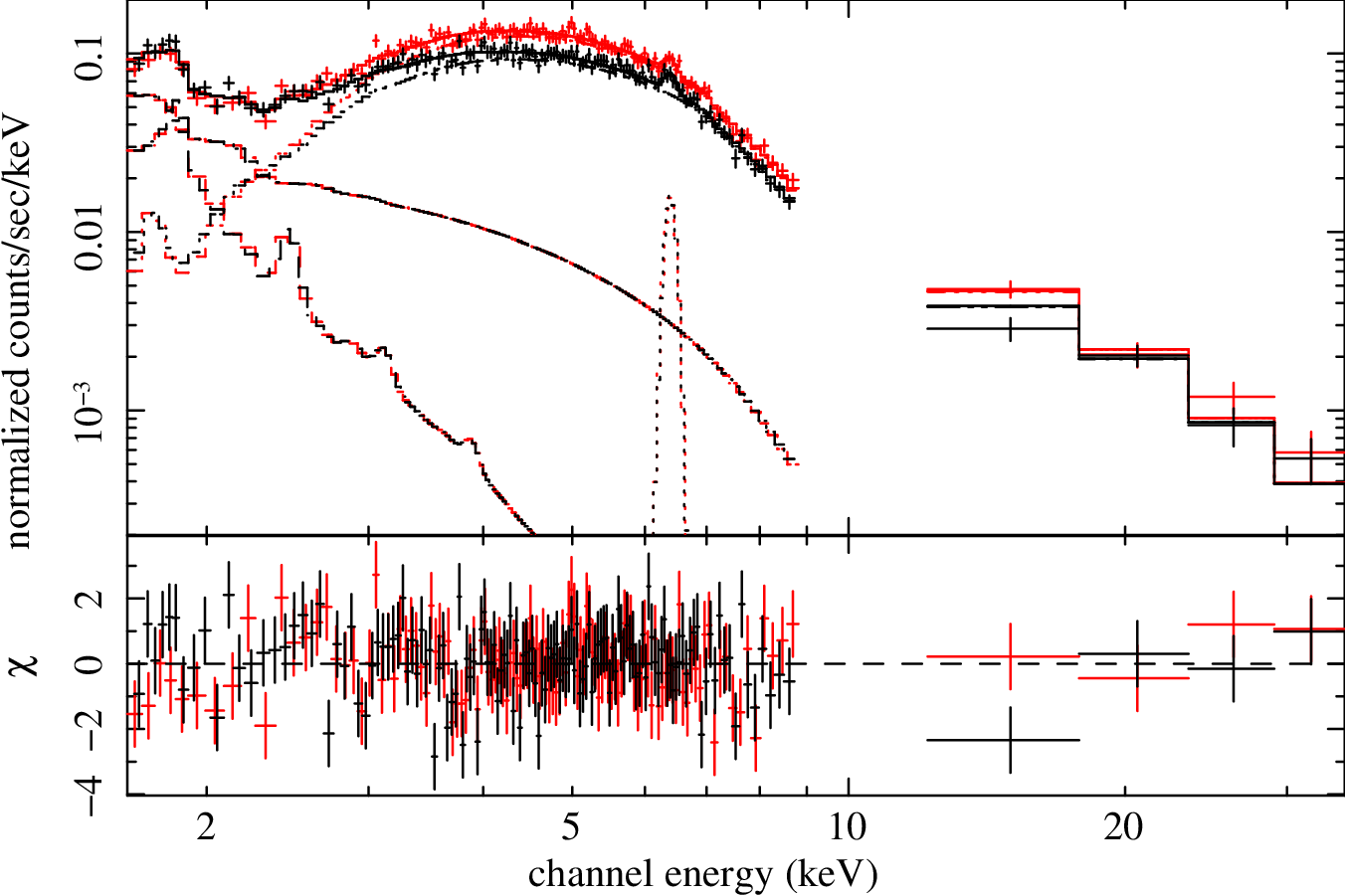}
  \end{center}
	\caption{The background subtracted spectra from
	the high-flux (red) and low-flux (black) phases,
	fitted with the same model as figure 7d.
	The XIS-BI spectrum, though employed in the fitting, is omitted from the plot for clarity.}
 \label{hl}
\end{figure}

To quantify the intensity-correlated spectral changes,
we fitted simultaneously the intensity-sorted spectra
with the same model as employed in section \ref{subsec:aba_ave}.
The normalization and temperature of the thermal plasma component
was fixed to those obtained with the time averaged spectra,
because such extended plasmas would not vary.
Furthermore, we constrained the bremsstrahlung normalization
and the Fe-K line flux to be the same between the two data sets.
This is because the former is thought to be an assembly of many faint point sources,
and the latter is considered to arise at too large distances to vary significantly on $\sim$ 100 ks.
We retained the constraint of $kT_{\rm br} =5$ keV.
The results became fully acceptable with $\chi^2/\nu = 435.8/436$, as presented in table \ref{hl_tbl} and figure \ref{hl}.
Thus, the nuclear emission became softer by
$\Delta \Gamma \sim 0.3$ as the intensity increased,
with a marginal evidence of an increase in $N_\mathrm{H}$.
This spectral steepening is most clearly visible in figure~\ref{n4258-hilo_ratio}
over the 5--9 keV range of the XIS data.
As a cross confirmation,
we tentatively constrained $\Gamma$ to be the same between the two data sets,
to find that the fit  worsens by $\Delta \chi^2 = 29.8$.
The fit degradation is smaller ($\Delta \chi^2 = 6.5$)
when we tie $N_\mathrm{H}$ instead of $\Gamma$.

In figure \ref{n4258-lc} and figure~\ref{n4258-hilo_ratio}, 
the 1--2 keV signals exhibit a hint of weak variation.
To examine this effect,
we tentatively allowed the bremsstrahlung normalization
to take separate values between the two phases,
while keeping  $kT_{\rm{br}}$ the same between them.
Then, the bremsstrahlung normalization became $\sim 10$\% lower
in the high-fllux phase than  in the other,
with a fit improvement by $\Delta \chi^2 =-4.6$ for $\Delta \nu = -1$.
The 2--20 keV bremsstrahlung luminosity was obtained
as $3.9 \pm 0.4$ and  $4.5 \pm 0.4$ in the high-flux and low-flux phases,
respectively, both in units of $10^{39}$ erg s$^{-1}$.
Therefore, the small (anti-phased) variation
seen in the 1--2 keV band of figure \ref{n4258-lc} 
is consistent with changes in the medium-hardness component.
Further examination of this issue is presented in section~\ref{section:discussion}.

The Fe-K line photon flux, when allowed to differ between the two phases, 
was obtained as $5.83 \pm 2.35$ and $5.95 \pm 1.82$
in units of $10^{-6} $ ph s$^{-1}$ cm$^{-2}$,
during the high-flux and low-flux phases, respectively,
together with $\Delta \chi^2 = -0.01$ ($\Delta \nu = -1$).
Therefore, our assumption of a constant Fe-K photon flux is self-consistent.
To examine an alternative case of constant EW,
we next constrained  the  Fe-K line flux in the low-flux phase
to be 0.8 times that in the figh-flux phase,
so that it becomes proportional to the continuum at 6.4 keV (figure~\ref{n4258-hilo_ratio}).
Then, the fit became rather worse by $\Delta \chi^2 = 2.7$.
Thus, the constant-line-flux hypothesis is more favored
than the constant-EW alternative,
though not conclusive.

Figure~\ref{diff_spec} shows `` difference spectra'',
obtained by subtracting the low-flux phase data from those in the high-flux phase.
As shown in table~\ref{hl_tbl} (last column),
the results can be described, in energies above 1.7 keV,
by a rather steep power-law with $\Gamma \sim 2.8$.
This reinforces the spectral softening in the high phase.
However, the value of $N_{\rm H}$ required by the difference spectra
is considerably higher than indicated by the time-averaged spectrum.
Actually, when $N_\mathrm{H}$ is fixed  at the value
($1.1 \times 10^{23}$ cm$^{-2}$) derived  with the time-averaged spectrum,
$\Gamma$ becomes $1.8 \pm 0.2$,
but the fit worsens by $\Delta \chi^2 = 10.0$.
While this could partially be due to a real increase in $N_{\rm H}$
toward higher fluxes as suggested by table~\ref{hl_tbl},
an artifact could also be working:
by subtracting a flatter/weaker power-law from a steeper/stronger one,
we will obtain a concave-shaped difference spectrum,
and its approximation by a single power-law would
require an artificially increased absorption.

\begin{figure}[t]
  \begin{center}
      \FigureFile(80mm,80mm) 
   {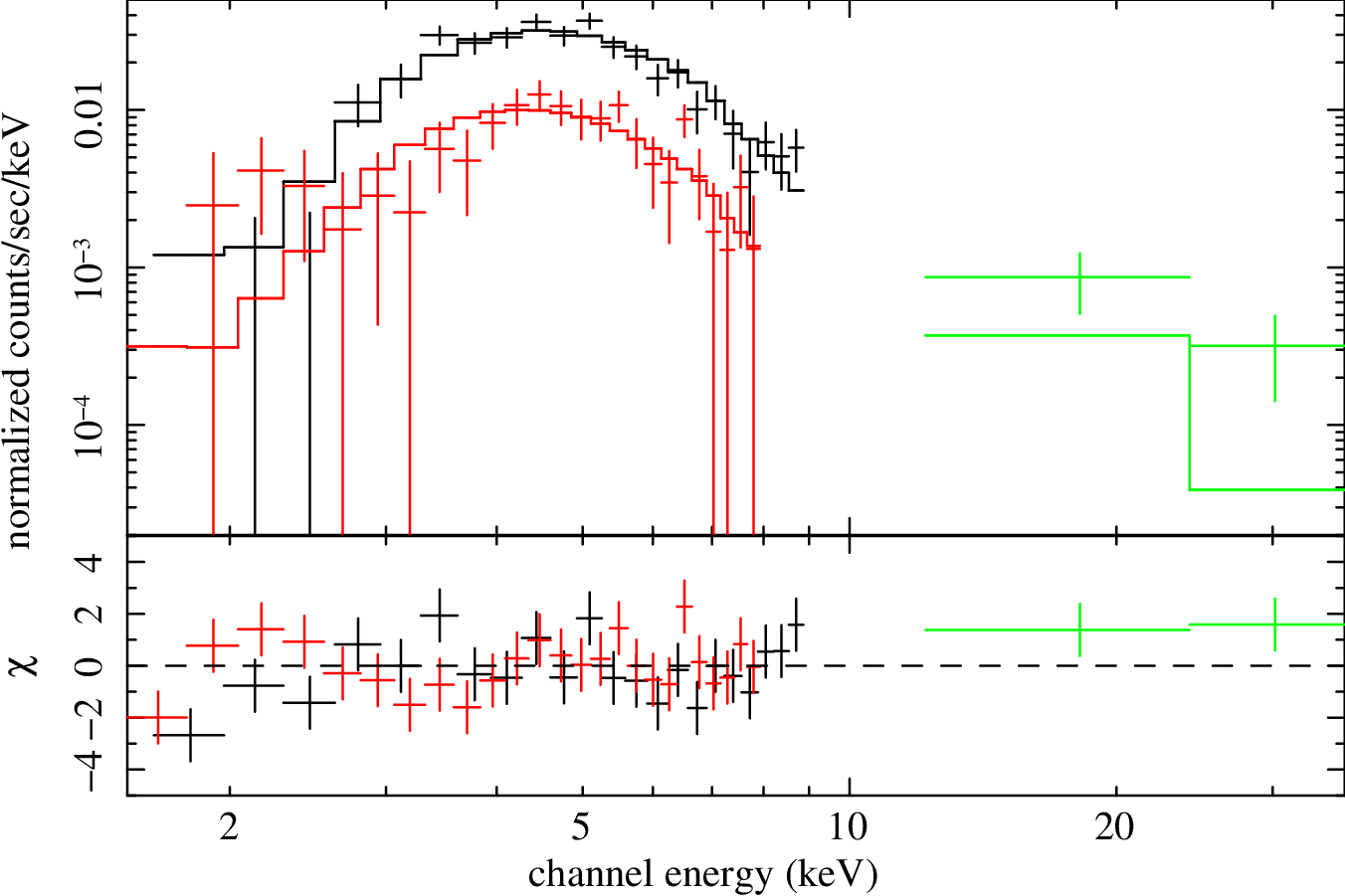}\\
  \end{center}
  \caption{The difference spectra taken with XIS-FI and HXD-PIN, fitted simultaneously 
  by an absorbed single powerlaw.
  The color specification is the same as in figure 7. 
 }\label{diff_spec}
\end{figure}

\section{Discussion}
\label{section:discussion}

\subsection{Summary of the results}

In 2006 June,
we observed NGC~4258  for  $\sim 100$ ks with  Suzaku.
In addition to the soft thermal plasma emission,
and a medium-hardness component presumably due to point sources,
the absorbed hard emission from its LLAGN was detected
with the XIS and HXD-PIN over a $\sim 2$ keV to $\sim 40$ keV band altogether.
The acquired wide-band spectra of the nucleus have been
described successfully in terms of a power-law with $\Gamma = 1.88$,
absorbed by $N_{\rm H} = 1.1 \times 10^{23}$ cm$^{-2}$.
The absorption-corrected 2--10 keV luminosity of the nucleus,
estimated as $\sim 8 \times 10^{40}$ erg s$^{-1}$,
is typical of this LLAGN (e.g., \cite{Max1994}, \cite{Reynolds2000}, \cite{Pietsch2002}, \cite{Yang2007}).
A narrow Fe-K line at 6.4 keV was detected
at  a relatively low EW of $\sim 50$ eV.
The reflection component was insignificant,
with an upper limit of  $f_{\rm refl}< 0.3$
if the reflector is assumed to have $i=60^\circ$,
although it is relaxed to  $f_{\rm refl}< 2.0$
if  the water-maser configuration of $i=83^\circ$ is adopted 
\citep{Miyoshi1995}.

On a time scale of  $\sim 100 $ ks,
we detected a clear hard X-ray variation by 20\%,
which can be explained by a change in the nuclear component.
As the hard X-ray flux increased,
the power-law component steepened by $\Delta \Gamma \sim 0.3$
with  marginal evidence for an increase in $N_{\rm H}$.
The varying component was described by
a relatively steep power-law with $\Gamma \sim 2.8$

\subsection{Implications of the time-averaged spectrum}

Being a prototypical LLAGN,
the most outstanding feature of the NGC~4258 nucleus
is of no doubt its very low X-ray luminosity,
both in the absolute and relative sense.
Hence, one of the most intriguing questions is
whether it exhibits any X-ray signatures (other than its low luminosity)
that are considered specific to LLAGNs.
When trying to answer this question,
the wide-band spectral coverage of Suzaku,
with its good energy resolution,  is considered essential.
As part of such attempts,
we may  quote X-ray photon indices of the intrinsic nuclear component
of  several representative Seyferts,
determined  with Suzaku through careful separation of the reflection component.
The results include  $\Gamma \sim$ 1.6 in NGC~4945 (\cite{Itoh2008}),
$\Gamma \sim 1.7$  in NGC~4388 (\cite{Shirai2008}),
$\Gamma \sim 1.8$  in MCG 5--23-16 (\cite{Reeves2007}),
and $\Gamma \sim$ 1.9  in NGC 3516 (\cite{Markowitz2008}).
Thus, our results on  NGC~4258,  $\Gamma \sim 1.9$,
does not distinguish this LLAGN from the other Seyferts.

Then, what do we find about circum-nuclear matter distribution in NGC~4258?
Generally, this can be probed using reprocessed signals,
including in particular the  fluorescence iron lines and the 
reflection component.
The measured EW ($\sim 50$ eV) of the former component,
which agrees with previous measurements from this LLAGN 
\citep{Max1994,Fiore2001},
is significantly smaller than those generally found among Seyfert galaxies
of both type 1 (typically $\sim 150$ eV) and type 2 (up to $\sim 1$ keV; \citet{Ueda2007}).
The latter component, i.e. reflection,  consists of three spectral features;
a deep Fe-K edge at $\sim 7$ keV, a Compton hump above 10 keV,
and a  hard continuum approximated by $\Gamma \sim -0.3$
(rising with the energy; figure~\ref{widebandspec}e) below the Fe-K edge.
While the last feature is confused in our spectrum with the 
point-source contribution,
the iron edge is clearly very weak (figures~5 and 6).
Furthermore, the HXD-PIN data, combined with the XIS flux points,
rule out the presence of a strong Compton hump.
Assuming a general configuration of $i=60^\circ$,
we have thus  obtained a tight constraint as $f_\mathrm{refl}< 0.3$,
which falls much below those typically found with Seyferts
($f_\mathrm{refl} \sim 1-2$; \cite{DeRosa2008}).

The above results on the Fe-K line and reflection both means
that the nucleus of NGC~4258 is devoid of thick materials with  large 
solid angles.
As a result, an inflated torus
(for which $i=60^\circ$ is regarded  as a reasonable approximation)
is unlikely to be present around it.
Nevertheless,  our direct view to the nucleus is obscured
by a thick material with $N_{\rm H} \sim 1\times 10^{23}$ cm$^{-2}$.
Then, the obscuring material must be localized
to a limited solid angle including our line of sight,
rather than forming a thick torus.
In fact, our data are consistent with a slab-like matter
distribution with  $\ensuremath{f_\mathrm{refl}} \sim 1$,
as long as it is viewed from an edge-on direction
(section~\ref{subsec:above1.7}).
One likely scenario is
that the  nearly edge-on and warped accretion disk,
which has been revealed by the water-maser results,
causes a grazing obscuration of the nucleus.

After NGC~4945 \citep{Itoh2008},
the NGC~4258 nucleus thus becomes a second example
in which the circum-nulear matter  is inferred to have a thin disk geometry
rather than a thick toroidal shape.
While such objects  are obviously considered relatively rare in a 
statistical sense,
we must investigate the reason why these objects lack a thick torus
that is usually considered to be a common attribute of an AGN.
The 2--10 keV luminosity of $8 \times 10^{40} $ erg s$^{-1}$,
measured from NGC~4258,
translates to a bolometric luminosity of $L_{\mathrm{bol}} \sim 6 
\times 10^{41}$,
if we use the argument by \citet{Marconi2004}, assuming that the multi-wavelength spectra of LLAGNs
are not much different from those of typical AGNs.
The estimated $L_{\mathrm{bol}}$ is still only $\sim 1 \times10^{-4}$
of the Eddington limit for an object of $3.6 \times 10^7~M_\odot$.
In contrast, the Eddington-normalized bolometric luminosity of NGC~4945
amounts to 0.3--1.0 (\cite{Itoh2008}).
Therefore, a low {\it normalized} luminosity is
unlikely to provide a common account for the lack of a torus.

When we  consider the absolute luminosity
(or mass accretion rate) instead of the normalized one,
not only NGC~4258 but also NGC~4945 has a considerably lower value
($7 \times 10^{42}$ erg s$^{-1}$ in 2--10 keV after correcting for the
absorption and Compton scattering; \cite{Itoh2008})
than typical AGNs (e.g., $10^{43-44}$ erg s$^{-1}$; \cite{Tueller2008}):
the high normalized luminosity of NGC~4945 is simply a results
of its low black-hole mass  ($1.4 \times 10^6~M_\odot$; \cite{Greenhill1997}).
Thus, a common feature of the two objects is very low accretion rates.
We hence speculate that the presence/absence of a thick torus
is related to the mass accretion rate rather than to the normalized luminosity.
Considering causality, one possible scenario  is
that a high accretion rate onto a giant black hole at the galaxy 
nucleus is realized
when the matter distribution has (for some unspecified reasons)
an inflated toroidal geometry,
whereas the rate remains low
when the matter has a thin disk configuration.

An accretion flow under an extremely low rate, like in the present case,
may be described by the ADAF model \citep{Narayan1995},
which assumes the optically-thick accretion disk
to be truncated at a large radius. 
\citet{Lasota1996} applied the ADAF prediction
to the X-ray spectrum of NGC~4258 measured by \citet{Max1994},
and argued that the innermost radius of the  accretion disk,
$r_{\rm{in}}$, should be larger than $\sim100~ r_{\rm{g}}$ in this LLAGN, 
where $r_{\rm{g}}$ is the Schwarzschild radius ($= GM/c^2$)
Since our spectrum is basically similar to that measured by \citet{Max1994},
the result by \citet{Lasota1996}  may also be applicable to our case.
Such a condition is consistent with our data,
because they also allow a solution with  $i=82^\circ$ and $\ensuremath{f_\mathrm{refl}} \sim 0$.
Incidentally, \citet{Max2008} measured $\ensuremath{f_\mathrm{refl}} \sim 0.4$ from Cygnus X-1,
together with $r_{\rm{in}} \sim 15~r_{\rm{g}}$.
Therefore, the concept of ADAF may be more applicable to
NGC~4258 than to Cygnus X-1.

\subsection{The time variations}

The time variation we detected,
on a time scale of  $\sim 100$ ks,
is one of the most rapid changes ever observed from NGC~4258.
We discovered  that  the power-law spectrum softens by $\Delta \Gamma \sim 0.3$
when the nucleus brightens up by 20\%.
This result is apparently reminiscent of
the positive correlation between $\Gamma$ and flux,
measured repeatedly from other AGNs;
e.g., from NGC~4151 with EXOSAT and Ginga \citep{Warwick1989,Yaqoob1992},
and from NGC~4051 with ASCA (\cite{Guainazzi1996}).
Such  positive $\Gamma$ vs. flux correlations allow two alternative 
interpretations.
One is to assume that this is an intrinsic property of the nuclear emission,
possibly related to cooling of  hot electron ``coronae"
that produce  hard X-rays via thermal Comptonization
of some soft seed photons (e.g., \cite{ST1979}).
The other is to consider the relation as an artifact,
produced by a superposition of a  power-law component
which varies without changing its slope,
and a constant (and harder) reflection component.

As to black hole binaries in so-called  Low-Hard state,
including Cygnus X-1 in particular,
Suzaku observations \citep{Max2008} revealed
that the former is actually occurring at least on time scales longer than 1 s.
(The latter mechanism does not work on this time scale,
since the reflection catches up with the intrinsic nucleus variation).
In the case of AGNs, in contrast, the latter is likely to be dominant,
because their ``difference" spectra can be generally
expressed by a single power-law,
of which the photon index is close to what is
found with the time-averaged spectrum
after separating the reflection component.
Such examples include NGC~4945 (\cite{Itoh2008})
and MCG 5--23-16 (\cite{Reeves2007}).
Returning to NGC~4258,
our results prefer the former interpretation,
because the reflection component is basically insignificant in this object.
Then, the mechanism responsible for the variation of this LLAGN
is suggested to be closer to that of black-hole binaries
in the Low-Hard state than to those of Seyfert galaxies.

An intriguing result is the weak variation in the 1--2 keV band,
apparently anti-correlated with that in the absorbed power-law.
One possible interpretation is
to invoke a variation of a small number of luminous point sources,
as suggested by the analysis conducted in section \ref{subsec:hl_ave}.
However, the implied luminosity change amounts to $\sim 5 \times 
10^{38}$ erg s$^{-1}$,
which would be too large to be explained by such a mechanism.
Furthermore, we would have to invoke a chance coincidence
to explain the apparent anti-correlation between the 1--2 keV and 
hard-band variations.
Therefore, it is more natural to consider
that the soft-band signals are contributed not only by point sources,
but also by the nucleus at some level,
and the variation therein is attributable to the nucleus contribution.
Since the absorbed power-law component falls far below ($\sim 1/30$)
the overall signal intensity at 1.5 keV (figure~\ref{ape_eeuf}),
this interpretation requires
that a small  (e.g., $\sim 1\%$) fraction of the intrinsic emission
reaches us without being absorbed.
Below, we construct a possible explanation based on this idea.

Let us  recall  that  the hard X-rays from Seyfert-like objtects are
likely to be emitted from a hot corona with a relatively large scale height,
via Compton scattering by thermal electrons \citep{Max2008}.
If such a corona is viewed from sideways,
its line-integrated Compton emissivity is expected to decrease rapidly
as a height away from the accretion plane
(like in the limb regions of solar coronae).
Then, the highest part of the corona,
which carries only a small fraction of the overall emissivity
(but geometrically rather extended),
could be rising above the grazing absorber and hence directly visible.
If  the height of this corona is  time variable,
the hard X-ray intensity will increase when the corona becomes less tall,
because the Compton optical depth of the corona will increase due to 
compression.
At the same time, the directly-visible  fraction of the coronal top 
region will decrease,
and will reduce the unabsorbed X-ray flux.
Furthermore, the absorber may have a gradient in the column density,
in such a way that it decreases away from the accretion plane.
Then, a more compressed corona will sample on average lower heights 
of the absorber,
leading to a slight enhancement  in  $N_{\rm H}$
as suggested by the data (table \ref{hl_tbl}).

As a consistency check of the above scenario,
a typical scale of such a corona is estimated as several tens $r_{\rm g}$
if an analogy to Cygnus X-1 is adopted \citep{Max2008}.
In the present case of NGC~4258,
such a corona can vary just on $\sim 50$ ks,
assuming a Keplerian time scale at such radii.
Furthermore, the Comptonizing corona of Cygnus X-1
was revealed to be highly inhomogeneous \citep{Max2008},
and  its hard X-ray intensity was suggested to increase
when the corona becomes ``less porous".
This is consistent with the height variation we invoked above,
because an inhomogeneous corona will naturally become less porous
when it becomes vertically shorter.
For a final speculation,
a corona may become vertically shorter and less porous
(thus leading to an increased hard X-ray intensity),
as its internal magnetic pressure is released by magnetic reconnection.

The authors would like to express their thanks to the Suzaku
team members. Our work was supported by Grant-in-Aid for JSPS Fellow.

{}

\end{document}